\documentclass[aps,prl,showpacs,twocolumn,superscriptaddress]{revtex4}
\usepackage{graphicx,amssymb,amsmath}

\begin{document}

\title{Observation of an unusual equilibrium in the molecular nano-magnet Fe$_8$}

\author{V. Villar}
\affiliation{Centre de Recherches sur les Tr\`es Basses Temp\'eratures, laboratoire associ\'e \`a l'Universit\'e Joseph Fourier, CNRS, BP 166, 38042 Grenoble cedex 9, France}
\author{E. Lhotel}
\affiliation{Centre de Recherches sur les Tr\`es Basses Temp\'eratures, laboratoire associ\'e \`a l'Universit\'e Joseph Fourier, CNRS, BP 166, 38042 Grenoble cedex 9, France}
\author{C. Sangregorio}
\affiliation{INSTM RU and Department of Chemistry, University of Firenze, Via della Lastruccia 3, 50019 Sesto Fiorentino, Italy}
\author{C. Paulsen}
\affiliation{Centre de Recherches sur les Tr\`es Basses Temp\'eratures, laboratoire associ\'e \`a l'Universit\'e Joseph Fourier, CNRS, BP 166, 38042 Grenoble cedex 9, France}

\begin{abstract}
Magnetization measurements made in small fields as a function of temperature reveal  an unusual equilibrium below 900~mK for the molecular nano-magnet Fe$_8$. Measurements of the relaxation of the magnetization demonstrate that the approach to the equilibrium is non-trivial and suggest that a competition exists between quantum tunneling of the giant spins and thermodynamic behavior.  As a result, at very low temperature the entropy of the spin system remains very large.
\end{abstract}

\maketitle

A great deal of interest has been devoted to the study of molecular nano-magnets \cite{SessoliACIE03}. One of the most prominent examples is the octanuclear iron cluster Fe$_8$,  formula $\{$[Fe$_8$(C$_6$H$_{15}$N$_3$)$_6$(OH)$_{12}$]Br$_7$H$_2$O$\}$Br$\cdot$8H$_2$0 which crystallizes in the triclininc P1 system. At low enough temperature, a  crystal of Fe$_8$ can be considered as an ensemble of identical, spin $S$~= 10 nano-magnets with an ising like anisotropy and an easy plain for the reversal of the spin. Spectacular effects have been observed such as temperature independent  quantum tunneling of the magnetization with non-exponential decay \cite{SangregorioPRL97},  measurements of the distribution of dipolar fields, hole digging in the distributions and the dependence of hyperfine fields \cite{OhmEPJB98,WernsdorferPRL99,WernsdorferPRL00}, and the observation of the Berry phase \cite{WernsdorferS99}.

The spin Hamiltonian for Fe$_8$ with  field applied along the easy axis is to a good approximation given by
\begin{equation}
\mathcal{H}=DS_z^2+E(S_x^2-S_y^2)+g\mu_BS_zH. \label{eq1}
\end{equation}

The last term is the Zeeman energy. The axial and in plane anisotropies, $D$~= -0.29~K and $E$~= -0.046~K have been measured by EPR spectroscopy \cite{BarraCEJ00} and other experimental techniques \cite{MukhinPRB01}. The Hamiltonian describes an energy level schema with $2S+1$~= 21 levels. The low lying levels $m$~= 10, 9 etc. are more or less distinct, but mixing do to the off diagonal terms smears out the upper levels $m<6$ \cite{BarraCEJ00}.  Above 2~K Fe$_8$ behaves as a superparamagnet  and the relaxation is dominated by thermal activation over an energy barrier given by $DS_z^2$~= 29~K. Below approximately  360~mK relaxation takes place by temperature independent resonant tunneling between the lowest lying $m$~= $\pm 10$ levels. The tunnel splitting for the ground state is very small $\Delta\simeq 10^{-7}$~K ($10^{-4}$~gauss)\cite{WernsdorferS99}. However  the resonance  is significantly broadened by internal dipole fields (width at half maximum is of the order of 400~gauss for the unpolarized state) \cite{OhmEPJB98,WernsdorferPRL99} and in addition relaxation by tunneling is aided by (rather slow) nuclear fluctuations which can bring spins that are near resonance into resonance \cite{ProkofevPRL98,TupitsynPRB04,FurukawaPRB00}. At intermediate temperatures thermally activated quantum tunneling occurs. However, the  tunnel splitting for $m$~= 6 or 5 is very large, which opens broad channels in the barrier. As the relaxation times for thermal  activation over the barrier become longer, tunneling through these channels becomes more likely which effectively reduces the height of the energy barrier. This in turn will reduce the susceptibility, and reduce relaxation times.

In this letter we report the observation of a plateau in the dc magnetization $M$ as a function of temperature  which  occurs  below approximately 0.9~K when measured in small fields. This simple observation has been more or less over looked in the literature, where most of the excitement has focused on the unusual effects at lower temperature. At first glance, it may seem natural to have a plateau when cooling. After all, relaxation times are increasing, and below 1~K become comparable to the characteristic times of most experiments. When cooling too fast, one turns the corner so to speak, and the system becomes frozen in a non-equilibrium state. We will demonstrate below, that the plateau is not an artifact of the cooling rate, and is in fact a new kind of equilibrium. 

All measurements were made using low temperature SQUID magnetometers developed at the CRTBT/CNRS. They are equipped with miniature dilution refrigerators that can reach temperatures  down to about 0.08~K. Absolute values of the susceptibility and magnetization  are made by the extraction method. A number of different single crystals (mass approximately 1~mg)  were measured. The samples were aligned with the easy axis along the field. 

\begin{figure}[h!bt]
\centering{\includegraphics[width=8cm]{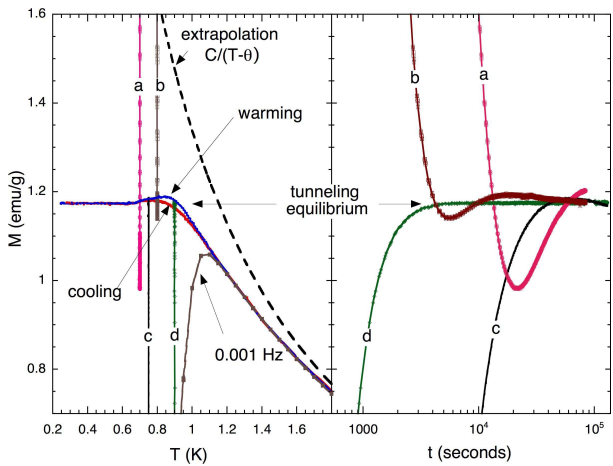}}
\caption{(left) M vs. T measured in a field of 25~Oe applied along the easy axis of a single crystal of Fe$_8$. The cooling rate was approximately 0.2~K/hr. Also shown is an ac magnetization curve for $f$~= 0.001~Hz. The dashed line is an extrapolation of the high temperature  Curie-Weiss behavior.  Below 0.9~K a plateau in the magnetization is observed which appears to be an equilibrium state. The vertical  lines show the relaxation of the magnetization toward the plateau (made at constant temperature and 25~Oe field) after first saturating the sample in a large positive field ($a $~= 0.7~K and $b$~= 0.8~K) or in a large negative field ($c$~= 0.75~K and $d$ = 0.9~K). (right) The same relaxation data vs. log time}
\label{fig1}
\end{figure}

Figure~\ref{fig1}a shows a plot of the field-cooled (and warmed)  dc magnetization  vs. temperature. The cooling and warm\-ing rates were  approximately  0.2~K/hour and the measurements were made in an applied field of 25~Oe. At these temperatures  25~Oe may be considered a small field because the magnetization is still linear with field and far from the saturation value  $M_{\rm sat}$ = 49.5~emu/gram  corresponding to 20~$\mu_B$ per molecule. Also shown is the real part of the ac magnetization measured at a frequency of 0.001~Hz. Below 0.9~K the magnetization becomes independent of the temperature and a plateau is observed. In addition, the value of  M/H on the plateau is nearly constant for $H<250$~Oe  , i.e. for fields within the first dipole broadened resonance.

The dashed line shown in the figure is an extrapolation of the high temperature  Curie-Weiss behavior and deserves some comment.  The extrapolation was made by plotting $1/\chi$  and using the restricted temperature range from approximately 2 to 6~K (a slight correction for demagnetization  effects was also made). In this temperature range the $1/\chi$ data fall on a straight line and is in good agreement with the $S$~= 10 ground state and an enhanced  susceptibility due to the rather large Ising like anisotropy of the Hamiltonian \cite{Ohmthesis}. The intercept of the $1/\chi$ plot gives a small positive Curie-Weiss  constant  $<$0.1~K, however the extrapolation is quite long, leading to large uncertainty ($\pm$~0.1) in this value. Nevertheless it does indicate that interactions if present, are weak when compared to 1~K. The departure from the Curie-Weiss behavior as the temperature is decreased below 2~K is due to growing importance of thermally activated  tunneling.

\begin{figure}[t]
\centering{\includegraphics[width=7cm]{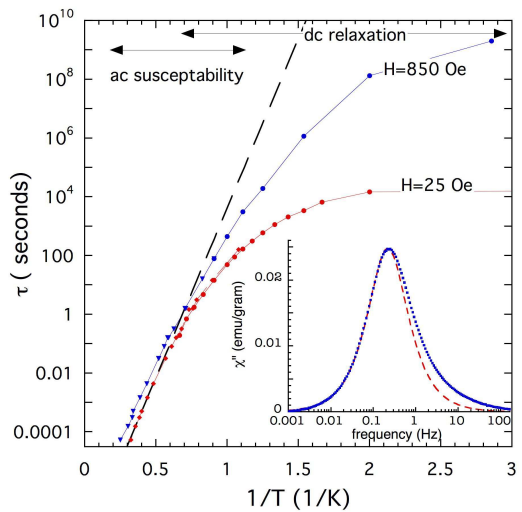}}
\caption{Relaxation times $\tau$ plotted against $1/T$ measured in the applied fields of 25~Oe (near  the first resonance) and 850~Oe (approximately between the first and second resonance). The dashed straight line is a fit to a thermal activation law using $E_b$~= 29~K from EPR data. The inset is a plot of  $\chi^"$ vs.  log frequency at constant temperature $T$~= 1.4~K. The dashed line is a lorentzian fit to the low frequency data points, i.e. data on the left of the peak.}
\label{fig2}
\end{figure}

It is well known that relaxation times $\tau$ for this super-paramagnetic system become very long as the temperature is decreased. In figure~\ref{fig2} a plot of $\ln\tau$ vs $1/T$ shows the temperature dependence of $\tau$ measured in two different applied magnetic fields: 25~Oe close to the first resonance, and 850~Oe which  is roughly between the first and second  resonance. A straight line on this plot would indicate that the relaxation follows a simple thermal activation  law $\tau$~= $\tau_0\exp E_b/kT$ over an energy barrier $E_b$. The dashed line is a fit using the high temperature  EPR value for $E_b$~= 29~K.

The low temperature data points ($T<1.4$~ K) of figure~\ref{fig2} 
were  made by fitting dc relaxation curves to a single exponential function \cite{note}. The  high temperature  data points of figure~\ref{fig2} ($T>0.9$~K ) have been determined from the peaks in the imaginary part of the ac susceptibility $\chi^"$. The inset shows a typical  plot of $\chi^"$ vs. the logarithm of the frequency made at fixed temperature $T$~= 1.4~K. Note that the curve is not symmetric but has a high frequency shoulder. The dotted line is a fit to the data using the low frequency side of the peak only and assuming the Casimir and du Pr\'e equation $\chi^"$$ \sim$ $\omega$$\tau$/1+$\omega^2$$\tau^2$.

The conclusions that can be drawn from figure~\ref{fig2}  and the above analysis is that at least experimentally, relaxation times for Fe$_8$ are well characterized. So is the appearance of the plateau below 0.9~K because the sample was  cooled too fast? Consider the following experimental observations which indicate that the plateau is an equilibrium state of the system:

\noindent 1.~The plateau is independent of the cooling rate over a very large range of rates. For example, cooling the sample from 1.5~K to 0.5~K at a relatively fast rate of 1~K per hour or cooling as slow as 1~K over 3 days results in the same plateau.

\noindent 2.~The sample can be slowly cooled to some temperature on the plateau,  for example 800~mK, and then monitored at this temperature for days. The magnetization remains constant. Note that at this temperature and for $H$~= 25~Oe, we estimate $\tau$ from figure~\ref{fig2} to be less than 6 hours, much shorter than our experimental time scale.

\noindent If the plateau is an equilibrium state, then it is interesting to see how the system moves toward this state when it is out of equilibrium. This can be done in the following ways;

\begin{figure}[b]
\centering{\includegraphics[width=7cm]{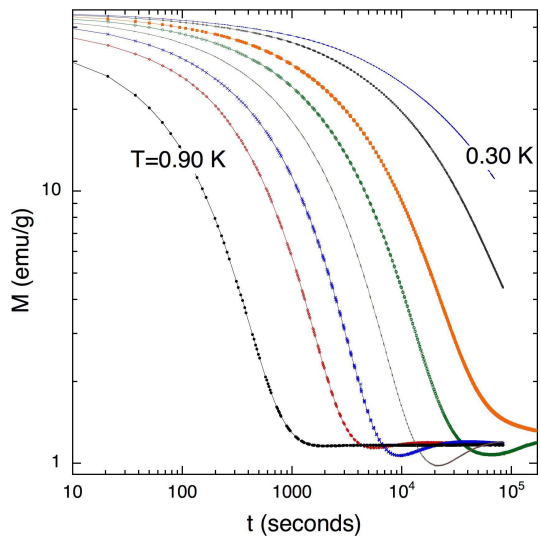}}
\caption{Relaxation of the magnetization vs. time measured in an applied field of 25~Oe and at various constant temperatures. (from left to right) $T$~= 0.9, 0.8, 0.75, 0.7, 0.65, 0.6, 0.5 and 0.3~K. The sample was first saturated in a high field.}
\label{fig3}
\end{figure}

\noindent 3.~By applying a field greater than a few teslas, the sample can be saturated.  Then at some fixed temperature below 0.9~K , the magnetic field can be abruptly changed to the value of the given field cooled magnetization, and the magnetization can be measured to see how the system relaxes from above, toward the plateau. Figure~\ref{fig1} (curves a and b) and figure~\ref{fig3} 
show some typical plots of the relaxation of the magnetization  in 25~Oe applied field and at various temperatures using this procedure.  As can be seen, the magnetization decreases, approaches and then actually overshoots the plateau, before reversing and approaching the plateau slowly. {\em The relaxation is non-monotonic!}

\noindent 4.~Alternatively, the sample can be saturated  in a large negative field. Then as above, the field can be rapidly changed to 25~Oe and the ensuing relaxation  recorded,  this time starting from below the plateau.  Two typical curves (c and d) are shown in figure~\ref{fig1}. The magnetization comes from below, relaxes upward, and overshoots the plateau, only to turn around, and approach the plateau slowly. Note however,  for this measuring field, the overshoot coming from below was less flagrant than from above.  A variation on this approach is to rapidly field cool the sample  ($<$~1 second) in the 25~Oe field to some temperature below 0.9~K  and watch how it relaxes toward the plateau from below.  Many curves have been measured using this technique and the results are essentially the same as above. 

\begin{figure}[h!tb]
\centering{\includegraphics[width=8cm]{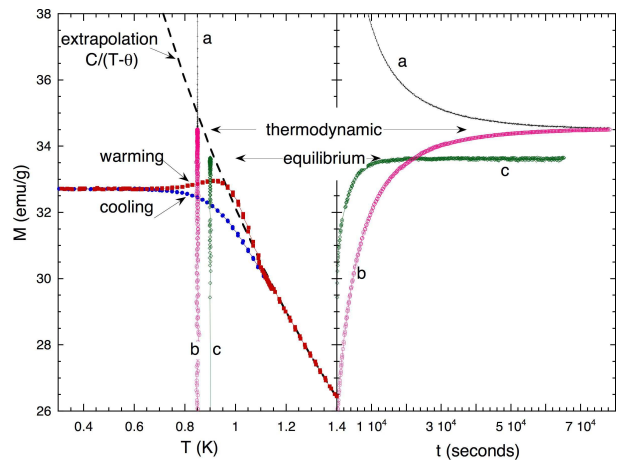}}
\caption{(left) Magnetization vs. temperature ``off resonance" for a cooling rate of approximately 0.4~K/hr and measured in a field of 850~Oe. The dashed line is an extrapolation of the high temperature Curie-Weiss behavior.  For this cooling rate a plateau in the magnetization is observed below 0.75~K.  The vertical  lines show the relaxation of the magnetization at constant temperature and in the same 850~Oe field after first saturating the sample in a large positive field ($a$~= 0.85~K) and after rapid field cooling the sample from $T>1.5$~K ($b$~= 0.85~K and $c$~= 0.9~K). Note that the  magnetization relaxes to a value much higher than that of the plateau and approaches the extrapolated Curie-Weiss value. (right) The same relaxation data plotted as a function of time showing the approach to equilibrium with no over-shoot.}
\label{fig4}
\end{figure}

\noindent  5.~The above effects can be destroyed and replaced by simple super-paramagnetic behavior  if tunneling between  the lowest lying states is suppressed. One way that this can be achieved is by applying a magnetic field along the easy axis that lies in between the first resonance at $H$~= 0 and the second resonance  ($M_z$~= -10 to $M_z$~= +9) at $H$~= 2000~Oe. Thus the left hand side of figure~\ref{fig4} shows the field cooled  (0.4~K/ hr) magnetization measured in an applied field of $H$~= 850~Oe. Note that this is a relatively large magnetic field at these temperatures and $M$ vs. $H$ is no longer linear.

As can be seen in the figure for this cooling rate a plateau in the magnetization is observed below 0.75~K. However this plateau depends strongly on the cooling rate and waiting times of the experiment. Thus a slower field cooling results in a higher plateau at lower temperature and we conclude that  it is not an equilibrium state of the system. This can be demonstrated by repeating the experimental procedures mentioned above. For example,  the vertical lines (a,b and c) in the left hand side of figure~\ref{fig4} are plots of the relaxation of the magnetization at constant temperature after first saturating in a large positive field (a) and after a rapid field cooling from 1.5~K (b and c). The same curves are also shown on the right hand side of the figure plotted as a function of time. The curves are monotonic, there is no overshoot and in fact can be well approximated to a single exponential relaxation. Most important however  is that after waiting a long enough time the curves converge to the extrapolated  low temperature thermodynamic magnetization.

From these experimental observations we conclude that the plateau, when measured in a small field as in figure~\ref{fig1}, represents an unusual equilibrium state of the system, and is clearly not the result one would expect from simple application of the partition function using Eq.~\ref{eq1}. The plateau could be the signature of the onset of an ordered state. However in our opinion, at least near  0.9~K, it results from the competition between thermal activation  which tends to order the system, and the disorder that results from ``indiscriminate" resonant quantum tunneling between the lowest lying states.  A simple outline of this process is as follows: in the small applied field of figures \ref{fig1} and \ref{fig3}, the magnetization value on the plateau $M=\mu_B(n_\uparrow-n_\downarrow$) represents about a 2.4\% excess in the populations of up spins $n_\uparrow$ with respect to down spins $n_\downarrow$ (considering the approximation of a two level system where the total number of spins is given by $n=n_\uparrow+n_\downarrow$). As the temperature decreases, down spins will tend to align with the field direction in order to reduce the free energy and give rise to super paramagnetic behavior $\chi$~= $C/T$. Now imagine a down spin that is in a positive local field $H_{\rm local}$. If $H_{\rm local}\gg\Delta$, then relaxation by resonant tunneling is ruled out. This spin can nevertheless flip and align with the local field by thermally  activated (non-resonant) tunneling. The temperature and field dependent relaxation time $\tau(T,H_{\rm local})$ will lie between  the two extremal curves given in figure \ref{fig2}. For example $10^{-3}<\tau<10^{-5}$~sec for $T$~= 0.8~K, well within our experimental waiting times. 

When this spin flips, it will change the local field of its neighbors, increasing the field on some, decreasing on others, but in particular it will move some spins into resonance or at least close enough to resonance for nuclear fluctuations to become important and thus enabling tunneling between the  $m$~= $\pm 10$ states.  However this tunneling does not discriminate between up and down spins (after all for these spins $|H|\sim\Delta\simeq 0$). On the other hand, statistically there will be more up spins than down spins, so the probability  that an up spin flips down is more likely, and thus the tunneling tends to decrease the magnetization  $M$. After a spin has tunneled, it will also change the local field of its neighbors, moving new spins on and off resonance. The process continues by a slow diffusion of tunneling spins through out the sample, and gives rise to a constant rearrangement of up and down spins.

In our model, as the temperature deceases, the entropy of the spins remains very large. For small applied fields as in our experiments, $n_\uparrow\sim n_\downarrow$,  and $S \lesssim Nk_B\ln 2$. This should be compared to superparamagnetic behavior  were spins align with the field and $S=0$ at $T=0$. Implicit in the calculation of the spin entropy is that  all configurations of the spin phase space are accessible and are ``visited'" during some characteristic time  $\tau$ which may be very long but finite.
As $T\rightarrow 0$, one might expect nuclear spins to order or at least freeze out. This would change significantly the tunneling dynamics by eliminating the nuclear fluctuations. Tunneling may continue, but at a reduced rate.

Recently, more experiments have been made which shed additional light on this curious behavior and will be reported elsewhere \cite{ccp}. In particular a transverse field was added to one of the magnetometers. By applying an appropriate transverse field along the hard axis tunneling can be greatly suppressed (a manifestation of the Berry phase for nano-magnets) \cite{WernsdorferS99,garg}. When tunneling is suppressed by this method, the results are similar to point 5: super-paramagnetic behavior is restored and the plateau depends on cooling rate. This further reinforces the idea that tunneling is responsible for the plateau.


\begin{thebibliography}{99}
\bibitem{SessoliACIE03}
R. Sessoli, D. Gatteschi, \textit{Angew. Chem. Int. Ed.} \textbf{42}, 268 (2003); G.~Christou, D.~Gatteschi, D.N.~Hendrickson, R.~Sessoli, \textit{MRS Bull.} \textbf{25}, 66 (2000).
\bibitem{SangregorioPRL97}
C. Sangregorio, T. Ohm, C. Paulsen, R. Sessoli, and D. Gatteschi, \textit{Phys. Rev. Lett.} \textbf{78}, 4645 (1997).
\bibitem{OhmEPJB98}
T. Ohm, C. Sangregorio, and C. Paulsen, \textit{Eur. Phys. J.}  B \textbf{6}, 195 (1998).
\bibitem{WernsdorferPRL99}
W. Wernsdorfer, T. Ohm, C. Sangregorio, R. Sessoli, D. Mailly, and C. Paulsen, \textit{Phys. Rev. Lett.} \textbf{82}, 3903 (1999).
\bibitem{WernsdorferPRL00}
W. Wernsdorfer, A. Caneschi, R. Sessoli, D. Gatteschi, A. Cornia, V. Villar, and C. Paulsen, \textit{Phys. Rev. Lett.} \textbf{84}, 2965 (2000).
\bibitem{WernsdorferS99}
W. Wernsdorfer and R. Sessoli, \textit{Science} \textbf{284}, 133 (1999).
\bibitem{BarraCEJ00}
A.-L. Barra, D. Gatteschi, and R. Sessoli, \textit{Chem. Eur. J.} \textbf{6}, 1608 (2000).
\bibitem{MukhinPRB01}
A. Mukhin, B. Gorshunov, M. Dressel, C. Sangregorio, D.~Gatteschi, \textit{Phys. Rev. B} \textbf{63}, 214411 (2001).
\bibitem{ProkofevPRL98}
N.V. Prokof'ev and P.C.E. Stamp, \textit{Phys. Rev. Lett.} \textbf{80}, 5794 (1998).
\bibitem{TupitsynPRB04}
I.S. Tupitsyn, P.C.E. Stamp, and N.V. Prokof'ev, \textit{Phys. Rev.} B \textbf{69}, 132406 (2004).
\bibitem{FurukawaPRB00}
Y. Furukawa, K. Kumagai, A. Lascialfari, S. Aldrovandi, F. Borsa, R. Sessoli, and D. Gatteschi, \textit{Phys. Rev.} B \textbf{64}, 094439 (2001).
\bibitem{Ohmthesis}
T. Ohm, Thesis, Univ. Joseph Fourier, Grenoble (1998).
\bibitem{note}
This is a rather poor approximation at very low temperature but becomes more reasonable near 1~K.  On the other hand single exponential fits for the relaxation curves measured in 850~Oe (although not perfect) are much better than those for 25~Oe over the entire temperature range.
\bibitem{ccp}
C. Paulsen and C. Sangregorio , \textit{submitted to LT24 August 2005, Orlando FL (AIP conf. proceedings)}.
\bibitem{garg}
A. Garg, \textit{Europhys. Lett.} B \textbf{22}, 205 (1993).
\end{thebibliography}
\end{document}